# Thermal Runaway, Flash Sintering and Asymmetrical Microstructural Development of ZnO and ZnO-Bi$_2$O$_3$ under Direct Currents

Yuanyao Zhang[1,2], Jae-Il Jung[2,3], and Jian Luo[1,2,*]

[1]Department of NanoEngineering; Program of Materials Science and Engineering, University of California, San Diego, La Jolla, CA 92093, USA

[2]Department of Materials Science and Engineering, Clemson University, Clemson, SC 29634, USA

[3]Interdisciplinary School of Green Energy, Ulsan National Institute of Science & Technology (UNIST), Ulsan 689–798, Korea

## Abstract

DC flash sintering of both pure and 0.5 mol. % Bi$_2$O$_3$-doped ZnO at a relatively high activating field of 300 V/cm has been investigated. It is demonstrated that even high-purity ZnO single crystals can "flash" at ~870 °C. In comparison, flash sintering occurs at a substantially lower onset temperature of ~550 °C in ZnO powder specimens, indicating the important roles of surfaces and/or grain boundaries. A model has been developed to forecast the thermal runaway conditions and the predictions are in excellent agreements with the observed onset flash temperatures, attesting that the flash starts as a thermal runaway in at least these ZnO based systems. Interestingly, enhanced grain growth is observed at the anode side of the pure ZnO specimens with an abrupt change in the grain sizes, indicating the occurrence of electric-potential-induced abnormal grain growth. With a large current density, the growth of aligned hexagonal single-crystalline rods toward the anode direction is evident in the ZnO powder specimen. Moreover, Bi$_2$O$_3$ doping defers the onset of flash sintering, which can be explained from the formation of space charges at grain boundaries, and it homogenizes the microstructure due to a liquid-phase sintering effect. The key scientific contributions of this study include the development of a model to predict the thermal runaway conditions that are coincident with the observed onset flash sintering temperatures, the clarification of how flash starts in ZnO based specimens, and the observation and explanation of diversifying phenomena of sintering and microstructural development under applied electric currents.

---

[*] Corresponding author. E-mail address: jluo@alum.mit.edu (J. Luo).

# 1. Introduction

Recently, field-assisted sintering technology (FAST), also known as spark plasmas sintering (SPS) and pulsed electric current sintering (PECS), has attracted great interest because it can achieve high sintered density at a lower temperature in a shorter time with an improved microstructure [1-6]. In 2010, Raj and colleagues invented "flash sintering" [6, 7], in which they used an electric field (20-100 V/cm) to initiate rapid densification of 3YSZ (3 mol. % $Y_2O_3$-stabilized $ZrO_2$) in just a few seconds at hundreds of degrees below the normal sintering temperatures. Subsequently, flashing sintering was demonstrated for 8YSZ (an ionic conductor) [8], $Co_2MnO_4$ (an electronic conductor) [9, 10], $La_{0.6}Sr_{0.4}Co_{0.2}Fe_{0.8}O_3$ (LSCF, a mixed ionic electronic conductor) [11], $SrTiO_3$ (a dielectric oxide) [12], MgO-doped $Al_2O_3$ (but not pure $Al_2O_3$) [13], $MnO_2$-doped $SnO_2$ (but not pure $SnO_2$) [14], $SiC-Al_2O_3-Y_2O_3$ (but not pure SiC or SiC-Al-$B_4$C-C) [15], $TiO_2$ [16], $Ce_{0.8}Gd_{0.2}O_{1.9}$ (GDC) [17], and $Y_2O_3$ [18]. An applied stress can further reduce the sintering temperature in flash-sinterforging [19]. Researchers also reported similar flash sintering (grain welding) effects induced by applied AC electric loadings using YSZ and $BaCe_{0.8}Gd_{0.2}O_{3-\delta}$ specimens, where they argued that the current flow though the specimens (instead of the electric field) is the key parameter [20-22].

Flash sintering differs from FAST because the applied electric field is typically higher, the sintering time is shorter, and the furnace temperature is lower. Thus, flash sintering can in principle be more cost-effective and energy-efficient. A mechanism for flash sintering must explain the simultaneous and discontinuous increases in mass transport kinetics and electrical conductivity. Originally, Raj *et al.* proposed three possible mechanisms [7]: flash sintering may be related to 1) Joule heating at grain boundaries that enhances grain boundary diffusion and electrical conductivity; 2) an avalanche nucleation of Frenkel pairs driven by the applied field; and/or 3) a non-linear interaction between intrinsic fields (space charges at grain boundaries) and the applied field that produces "a catastrophic change in self-diffusion at grain boundaries" [7]. More recent studies attributed the rapid sintering to a combination of Joule heating and defect generation (including possibly unconventional avalanches of Frenkel defects) [6, 13, 18, 23] or enhanced ionic and electronic transport long selectively-heated (and even selectively-melted) grain boundaries and dislocations [24]; it was argued that that Joule heating alone is not sufficient for accounting for the observed fast densification [6, 13, 18, 23]. Chen and co-workers proposed electro-sintering of 8YSZ is due to ionomigration of pores resulted from surface diffusion of cations [25-27]. In this study, we proposed a model to predict the thermal runaway temperature that is coincidental with the observed onset flash temperature, thereby attesting that the flash starts as a case of thermal runaway in ZnO.

Furthermore, recent "two-electrode experiments" revealed interesting and intriguing observations of the field effects on grain growth [6]. Conrad and colleagues showed that a relatively weak applied DC or AC field could inhibit grain growth of 3YSZ significantly [28-32]. Consequently, sintering is enhanced because a smaller grain size provides a greater driving force. In a separate controlled experiment, Raj and colleagues also suggested that an applied modest DC field of ~4 V/cm can inhibit the grain growth in 3YSZ [33]. Two mechanisms have been proposed: Conrad explained this field phenomenon by the reduction in the grain boundary energy through interactions of the applied fields with the space charges near grain boundaries [30, 31]. An alternative explanation was that Joule heating at grain boundaries raised the local temperature and reduces grain boundary energy by an entropic effect; this not only reduced the driving force, but also created a pinning effect [33].

More recently, Chen and colleagues demonstrated that an applied electric current (of ~50 $A/cm^2$) could enhance the grain boundary mobility (by >10 times) in the cathode side discontinuously in 8YSZ, leading to abnormal grain growth [34]. They attributed this effect to the accumulation of supersaturated oxygen vacancies on the cathode side that caused cation reduction to lower its migration barrier [34]. In this study, we have observed a somewhat opposite effect: discontinuous (abnormal) grain growth and/or coarsening in the anode side in ZnO during the flash sintering, which can be explained from the possible occurrence of an electric-potential-induced grain boundary (defect) structural transition by extending and combining Chen and colleagues' concept (discussed above) [34], Tuller's theory of grain boundary defect chemistry in ZnO [35], and the idea of grain boundary complexion transitions [36].

Most recently, flash sintering of nanocrystalline pure ZnO under AC fields between 0 and 160 V/cm was reported, where normal grain growth was observed [37]. In this study, we applied DC currents at a higher field of 300 V/cm to high-purity ZnO single crystals as well as pure and 0.5 mol. % $Bi_2O_3$-doped ZnO powder specimens. We have observed a number of interesting and intriguing phenomena, including the flash (thermal runaway) of ZnO single crystals, reduction of the flash temperature in powder specimens, anode-side abnormal grain growth (in contrast of the cathode-side abnormal grain growth reported for 8YSZ [34]), growth of aligned single-crystalline rods, and doping effects on deferring the onset flash sintering and homogenizing microstructures, which have greatly deepened and enriched our fundamental understanding of the sintering and microstructural development under electric currents. Moreover, a quantitative model has been developed for predicting the thermal runaway conditions that are in excellent corroboration with experimentally-observed onset flash temperatures, thereby suggesting that the flash starts as a thermal runaway in at least ZnO based specimens.

## 2. Experimental

The pure and 0.5 mol. % $Bi_2O_3$-doped powder samples were fabricated by using purchased ZnO (Sigma Aldrich, St. Louis, Missouri, USA: >99.9% purity, <0.5 μm particle size) and $Bi_2O_3$ (Sigma Aldrich: ≥99.8% purity, 90-210 nm particle size) powders. Nominally pure ZnO powders were ball milled with alumina media for 2 hours in isopropyl alcohol with 0.5 wt. % of binder (10 wt. % of polyvinyl alcohol or PVA dissolved in isopropyl alcohol). 0.5 mol. % $Bi_2O_3$-doped ZnO powders were prepared by ball milling mixtures of the oxide powders and 0.5 wt. % of binder in isopropyl alcohol for 10 hours. All powders were subsequently dried in an oven chamber at 80 °C for 12 hours after milling. The 0.5 mol. % $Bi_2O_3$-doped ZnO powders were calcined at 600 °C for two hours in air in a covered Pt crucible, followed by another round of 2-hour ball milling in isopropyl alcohol with 0.5 wt. % of binder and subsequent drying. Dried pure and 0.5 mol. % $Bi_2O_3$-doped ZnO powder cakes were pulverized in a mortar and sieved under 150 mesh of sieve. The resultant granulated powders were uniaxially pressed at ~300 MPa in a mold (¼ inch diameter) into green specimens (disks) with the approximate dimensions: *D* (diameter) = 6.4 mm and *H* (thickness) = 4 mm. The average bulk densities (± one standard deviations) of pure and 0.5 mol. % $Bi_2O_3$-doped ZnO green specimens were 63.5 ± 0.8 % and 65.0 ± 2.2 %, respectively, of the theoretical densities. The green specimens were then heated at the ramping rate of 5 °C per minute to 500 °C and baked at 500 °C isothermally for one hour to burn out the binders. After measuring the dimensions and weight of each sample, both sides of the specimen were pasted by Pt inks with Pt wire buried underneath. The pasted specimens were then heated at the ramping rate of 20 °C per minute and baked at 500 °C for 20 minutes. This baking process was repeated approximately eight times until the Pt wire was connected solidly to the dried Pt pastes.

The electrodized sample was placed close to the thermocouple within a tube furnace, with Pt wire connected to the power source. Each specimen was heated at the ramping rate of 5 °C per minute under an (initial) electric field at 300 V/cm (calculated based on the initial specimen thickness). Similar to a typical flash sintering experiment, the applied voltage was kept a constant until the resultant current reach a preset maximum value ($I_{max}$ = 1 and 4 A, respectively, for this study), at which point the power source switched from the voltage-control mode to a current-limited mode. The maximum current densities and final electric fields in the activated state were estimated based on the actual measured dimensions of sintered specimens and listed in Table I, along with other experimental conditions and results. When the power density reached the maximum, the electric power source was kept on for an additional 30 seconds, with furnace being turned off (the furnace temperature did not change significantly with the 30 seconds). Then, the electric power was switched off and the specimen was cooled down within the furnace. The weight and dimension of the sintered specimens were measured to obtain the final bulk density.

High-purity ZnO single crystals (>99.99% purity, 5 mm × 5 mm × 0.5 mm, double-side polished) were purchased from MTI Corporation (Richmond, California, USA). Platinum was sputtered on the both sides of the single crystal specimens using a Denton Discovery 18 Sputter. The surrounding areas (sides of the single crystals) were slightly grounded by SiC papers after sputtering. The specimen was placed in a horizontal tube furnace and attached with Pt wires on both sides to apply a 300 V/cm electric field to conduct a benchmark experiment with identical heating scheme as the powder specimens. The maximum allowed currents ($I_{max}$) were again set to be 1 and 4 A, respectively, although the actual current only reached ~1.45 A in the latter case.

To further test our model with high precision, we performed a second experiment on a new ZnO powder specimen with two improvements. First, a high-purity ZnO powder (Sigma Aldrich, St. Louis, Missouri, USA: >99.99% purity; grain/particle size of the green specimen: 120 ± 50 nm) was used. Second, a high-resolution ammeter (a Prova 11 AC/DC mA Clamp Meter, Sigma Aldrich, St. Louis, Missouri, USA) was employed to allow us to measure more data points prior to the flash event with higher accuracies to test our thermal runaway model more critically and accurately. The rest of experimental procedures were identical to those used for the flash sintering experiments of the single-crystal specimens.

Scanning electron microscopy (SEM) was carried out using a field-emission microscope (Hitachi SU6600, Japan) to characterize the microstructures. Grain sizes were measured using a standard intercept method by drawing 30 lines in the vertical, horizontal and diagonal directions of the SEM micrographs (excluding the lengths of intercept sections with voids) and assuming a geometric factor 1.5 (for ideal spherical grains).

## 3. A Model to Predict Thermal Runaway

We first present a model to predict the occurrence of a thermal (coupled with electric) runaway. We wish to point out that there are two fundamental scientific questions about the underlying mechanisms of the flash sintering, *i.e.*,

(1) How does the "flash" start?

(2) How do densification and sintering occur after the start of the flash?

The current modeling approach aims to address only the first (but not the second) question, which is an important open scientific question by itself.

We want to further emphasize that thermal runaway is one (but perhaps not the only) possible mechanism that a flash can start. Yet, in the next section, we will show that the predicted thermal runaway temperatures are coincidental with observed onset flash temperatures for all ZnO based powder and single-crystal specimens studied here, thereby suggesting that the flash indeed starts as thermal runaway in at least these ZnO based specimens.

For a specimen under an applied electric field in a furnace, the rise of the specimen temperature is determined by the energy conservation law:

$$\sigma(T_S) \cdot E^2 \cdot V_S = \dot{Q}(T_S, T_F), \tag{1}$$

where $T_S$ and $T_F$ are the specimen (S) and furnace (F) temperatures, respectively, $E$ is the applied electric field, and $V_S$ is the volume of the specimen. $\dot{Q}(T_S, T_F)$ is the rate of heat dissipation (including contributions from heat conduction, convection and radiation) that depends the particular geometrical configuration and heat transfer environment. $\sigma(T)$ is the electric conductivity, which should include both the normal temperature-dependent conductivity and any contribution from non-equilibrium defects generated by the flash process (if there is any).

The heat dissipation should vanish when $\Delta T \equiv T_S - T_F = 0$, while the left side of Eq. (1) is always positive. Thus, the specimen temperature will rise. Moreover, if a positive and finite solution of $T_F$ ($\Delta T$) exists for Eq. (1), the temperature rise is stable. An unstable temperature rise (*a.k.a.* a thermal runaway) may occur under the condition:

$$E^2 V_S \left. \frac{d\sigma}{dT} \right|_{T_S} > \alpha, \tag{2}$$

where

$$\alpha \equiv \frac{\partial \dot{Q}(T_S, T_F)}{\partial T_S} \tag{3}$$

is a general parameter that characterizes the increase of heat transfer (dissipation) rate with the increasing specimen temperature, and it can be quantified for heat conduction, convention, and radiation, if all heat transfer parameters are known. For a case of blackbody radiation, a simple analytic equation exists:

$$\alpha = 4\sigma_{\text{Stefan}} T_S^3 A_S, \tag{4}$$

where $\sigma_{\text{Stefan}}$ (= 5.67 × 10$^{-8}$ W/m$^2$K$^4$) is the Stefan-Boltzmann constant, and $A_S$ is the surface area of the specimen. It is interesting to note that $\alpha$ only depends on the specimen temperature ($T_S$) for this case (Eq. (4)), as well as cases where simple linear laws ($\dot{Q}(T_S, T_F) = K(T_S - T_F)$) for the heat conduction and convention are applicable and $K$ can be assumed only depend on $T_S$ (a valid approximation for small $\Delta T$). This allows Eq. (2) to be solved numerically by a graphic construction method, which will be described and implemented in the following section.

The thermal runaway condition (Eq. (2)) implies that an unstable arise of temperature will occur if more heat is generated than it can be dissipated with an increase in the specimen temperature. In such a case, a temperature rise will lead to an increase in conductivity, which will subsequently increase electric power dissipation/heat generation and heat the specimen further in a positive feedback loop, which will essentially leads to a thermal runaway. This thermal runaway condition can be rewritten as:

$$\left.\frac{d\sigma}{dT}\right|_{T_S} > \frac{\alpha}{E^2 V_S}, \tag{5}$$

where the left side is a material property (which also depends on the microstructure) and the right side depends on the specimen geometry and other experimental conditions (the applied field and the specific heat dissipation environment).

In the next section, we will demonstrate that the above model can predict the thermal runaway temperatures of both ZnO single crystals and powder specimens that are coincidental with the observed onset flash temperatures in all cases. The excellent agreements between the predictions and experiments ascertain the underlying hypothesis, *i.e.*, a flash event starts as a simple case of thermal runaway (without the need of avalanching non-equilibrium defects as previously assumed) at least for the case of ZnO; however, we emphasize that this does not exclude the possible formation of non-equilibrium defects after the occurrence of the flash/thermal runaway in the activated state to accelerate sintering (to achieve fast densification rates that do not appear to be possible from simple extrapolations of temperature-dependent sintering rates, as prior studies have demonstrated [6, 13, 18, 23]) and affect microstructural development.

Specifically, we will use the experimentally-measured $\sigma(T)$ to predict the thermal runaway conditions and to compare them with the experimentally-measured onset flash temperatures in the next section. The excellent agreement between the predicted thermal runaway conditions and observed onset flash temperatures is a strong evidence to support that the flash starts as a thermal runaway in these systems.

We acknowledge that the current modelling approach aims to explain how flash starts, but does not provide mechanistic understanding of the densification during the flash sintering. In addition, the thermal runaway model presented here is phenomenological in term that it works regardless whether the underlying conductivity $\sigma(T)$ is electronic or ionic. Nonetheless, the predictions are genuine without using of any free parameters (despite that we fit experimentally-measured $\sigma(T)$ to Arrhenius functions in the current cases of ZnO; we recognize that other $\sigma(T)$ functions may be needed for materials with non-Arrhenius behaviors).

Finally, we noticed a study published by Todd *et al.* in early 2015 [38] after the initial submission of this manuscript, where a similar thermal runaway model was proposed (independently, with a slightly later submission date) to explain the occurrence of flash in 3YSZ (the most extensively studied flash sintering systems). Todd *et al.* [38]'s SU (static uniform) model is equivalent to the thermal runaway model presented here, although they formulated the criteria using slightly different equations and plot the "heating/cooling" diagrams differently. Todd *et al.* [38] also proposed a DNU (dynamic, non-uniform) model to explain the transit behaviors, particularly the incubation time, with successes. A key difference between the two studies is the 3YSZ is an ionic conductor and ZnO is an electronic conductor (a semiconductor self-doped with Zn interstitials and/or oxygen vacancies). The facts that similar thermal runaway models can be used to predict the onset of flash in both cases (as discussed in more details in the next section) demonstrates the general applicability of the thermal runaway mechanism for the start of flash beyond one system.

## 4. Results and Discussion

### 4.1. Flash of ZnO Single Crystals

A main character of flash sintering is represented by abrupt and simultaneous increases in both the specimen temperature and the electric current. To establish a baseline, we measured the electric current and voltage as functions of furnace temperature and calculated the electric power/heat generation rates and conductivities for two ZnO single crystals with an (initial) applied electric field of 300 V/cm. Fig. 1 shows the nominal electric powder density (= the volumetric heat generation rate; assuming no changes in the specimen volume for simplicity) *vs.* furnace temperature curves. For high-purity ZnO single crystals, the electric power density increased gradually and reached ~0.1W/mm$^3$ at ~870 °C, which should correspond to a moderate rise of the specimen temperature. A thermal runaway or "flash" occurred at $T_F$ = 870 °C and $T_F$ = 877 °C, respectively, for the two ZnO single crystals (Fig. 1; Table I).

In the current experiments, it was difficult to directly measure the specimen temperature. Thus, we adopted an approach proposed by Raj *et al.* to estimate the specimen temperatures by the black body radiation model [23]:

$$T_S^4 - T_F^4 = \frac{W}{\sigma_{\text{Stefan}} A_S}, \tag{6}$$

where $W$ is the input electric power that was obtained experimentally. Subsequently, we plotted the measured conductivity (in a logarithmical scale) *vs.* the reciprocal of the absolute specimen temperature curves in Fig. 2. While we recognize that this approach of estimating specimen temperature ignores the contributions of heat conduction and convection and the specimen is hardly an ideal blackbody (and, unfortunately, we do not have the available parameters to do better estimations), a good Arrhenius relation (good linearity in Fig. 2) has been obtained for the data obtained from the single crystals before the occurrence of the flash events. This provides us some confidence on the estimated specimen temperatures and it further allows us to fit the conductivity *vs.* temperature curves to an empirical Arrhenius equation for ZnO specimens (but not for some other ceramic materials, where the underlying conduction mechanisms are complex and the temperature-dependent conductivity is non-Arrhenius):

$$\sigma(T) = \sigma_0 \cdot e^{-\frac{h}{kT}}, \tag{7}$$

where $k$ is the Boltzmann constant and $\sigma_0$ is a pre-exponential constant. The activation enthalpy, $h$, was fitted to be 1.98 ± 0.05 eV (190 ± 5 kJ/mol) and 1.89 ± 0.03 eV (182 ± 3 kJ/mol), respectively, for two single-crystal specimens (by excluding the first two data points, where the measurements were less accurate because of the low currents, and data points after the flash events, where the estimated specimen temperatures might have relatively large errors).

It is important to note that the Arrhenius equation also applies well for the ZnO powder specimens, as demonstrated in Fig. 2(b). Thus, Eq. (7) is adopted here for ZnO specimens in this study (as well as Todd *et al.*'s study [38] of 3YSZ), while we fully recognize that Eq. (7) or a simple Arrhenius relation does not apply to many other ceramic materials.

In Fig. 2, we also plotted the measured conductivities *vs.* estimated specimen temperatures for the activated state (after the occurrence of the flash events). For ZnO single crystals, the measured conductivities are lower than what would be extrapolated from the Arrhenius equation. This implies the absence of the abnormally-high conductivities due to non-equilibrium defects (such as an abnormal avalanche of Frenkel defects) in ZnO; however, it does not exclude the possible generation of some non-equilibrium defects in the activated state. It should be noted that estimated specimen temperatures could have relatively larger errors for the specimens in the activated state (with great $\Delta T$), which may account for the lower measured conductivities.

Using the fitted temperature-dependent conductivity relation, we calculated the differentiate heat generation rates (the left side of Eq. (2)) *vs.* specimen temperature curves for two ZnO single crystals and plotted them in Fig. 3(a). In the same figure, we also plotted the differentiate heat dissipation rate ($\alpha$) *vs.* specimen temperature curve that was calculated based on the specific specimen geometry and the applied electric field of 300 V/cm, assuming the black body radiation model (using Eq. (4)). The intersection of the curves of differentiate heat generation and dissipation rates represents the solutions of Eq. (2), or the flash condition, above which more heat is generated than that can be dissipated, leading to an unstable temperature rise (thermal runaway) or flash.

Specifically for the two ZnO single crystals, the predicted specimen temperatures for a thermal runaway (presumably the onset of the flash) are $T_S = 940$ °C and $T_S = 952$ °C, respectively, whereas the estimated specimen temperatures right before the flash from experiments are $T_S = 962$ °C and $T_S = 967$ °C (Table I), respectively, which are only 15-22 °C higher, representing good agreements between the model and experiments. Furthermore, the furnace temperatures for the thermal runaway can be predicted from Eq. (6) by using the predicted $T_S$ and the electric power dissipation computed from the fitted $\sigma(T)$; the predicted furnace temperatures for thermal runaway are $T_F = 870$ °C and $T_F = 877$ °C, respectively, which are virtually identical to the experimentally-observed onset flash temperatures (Table I) with <1 °C errors!

While such surprisingly-good agreements may be somewhat coincidental, we want to point out that errors for predicted furnace temperatures ($T_F$) should be less than those for the predicted specimen temperatures ($T_S$) under the condition that an accurate $\sigma(T)$ function is used (because the errors caused by ignoring the heat conduction and convection, which are the major errors for the current model, can be partially canceled in estimating $T_S$). Specifically, in the current case, an accurate fit of $\sigma(T)$ is possible because of the good linearity in the log($\sigma$) *vs.* 1/$T$ curves before the flash. This is because the main error for the current calculation comes from the deferential heat dissipation rate ($\alpha$). Since $T_F = T_S - \Delta T$ and an error in $\alpha$ would change in the estimated $T_S$ and $\Delta T$ in the same direction; thus, the error for predicting $T_F$ can be less.

In summary, excellent agreements of the predicted thermal runaway temperatures and the observed onset flash temperatures have been achieved for single-crystal specimens (Fig. 3(a)), which has ascertained the proposed model and the underlying hypothesis that the flash starts as a thermal runaway.

### 4.2. Flash Sintering of ZnO Powder Specimens

We observed that flash sintering started at $T_F$ = 565 °C and $T_F$ = 553 °C, respectively, for two ZnO powder specimens (Fig. 1; Table I). Since the most recent study of AC flash sintering of ZnO found the onset flash temperatures to be at ~670 °C under 80 V/cm and ~620 °C under 160 V/cm, respectively [37], the current observation of onset flash sintering at ~553-565 °C under a 300 V/cm DC (instead of AC) field shows a consistent trend of decreasing onset flash sintering temperature with the increasing applied field, regardless whether a AC or DC field is applied (although the results presented in the next section will show that DC and AC fields can have very different effects on grain growth and microstructural development).

This study demonstrates that the onset flash sintering temperatures of powder specimens ($T_F$ = 550-600 °C) are more than 300 °C lower than the flash temperatures of ZnO single crystals ($T_F$ = 870-877 °C) under the same applied DC electric field of 300 V/cm. This comparison suggests the important role of free surfaces (and/or grain boundaries) in initiating and sustaining the flash sintering. Specifically, it is well established that ZnO often contains surface conduction layers with high concentrations of free electrons (on the order of $10^{12}$ electrons per $cm^2$) [39]; thus, it is possible that the flash in ZnO powder specimens can occur via enhanced electric conduction along the free surfaces of ZnO particles. Based on Arrhenius extrapolations, the apparent conductivities of ZnO powder specimens are one order of magnitude higher than those of single crystals at the same temperatures (Fig. 2), which is consistent with the hypothesis that enhanced conduction along free surfaces reduces the onset of flash sintering.

Using the (only) two measured conductivity data for first two ZnO powder specimens before the flash (Fig. 2(a)), we fitted the $\sigma(T)$ to Eq. (7) and calculated the differentiate heat generation and dissipation rates to predict the flash conditions via the same method used for single crystals. The results are shown in Fig. 3(b) as the Specimens #1 and #2, where the predicted and experimentally-estimated specimen temperatures at the thermal runaway/onset flash sintering are $T_S^{(predicted)}$ = 650-662 °C and $T_S^{(exp.)}$ = 675-680 °C, respectively; the corresponding predicted and measured furnace temperatures are $T_F^{(predicted)}$ = 524-545 °C and $T_F^{(exp.)}$ = 553-565 °C, respectively. The agreements (<30 °C in all cases) are satisfactory, given the relative large errors for measuring conductivities at low currents (that are measured using a low-resolution ammeter).

A major error source for predicting the thermal runaway conditions for ZnO powder Specimens #1 and #2 in Fig. 3(b) came from the fact that we could measure only two data points before the flash for the powder specimens due to the low resolution of our prior ammeter (Fig. 1). Thus, we further conducted an additional experiment using a new high-resolution ammeter that allowed us to measure many data points before the occurrence of the flash. These new data are represented by the open squares in Fig. 1 and Fig. 2 (referred to as the "new ZnO powder specimen") and the Specimen #3 in Fig. 3(b). There are three useful observations. First, this new experiment used a higher-purity ZnO (99.99%) powder, which presumably led to a higher onset flash temperature (because the impurities in the other 99.9% ZnO can serve as dopants that increase the conductivity and lower the onset of the thermal runaway). Second, this new experiment clearly showed the conductivity of ZnO powder specimen follows the Arrhenius relation (Eq. (7)), as shown in Fig. 2(b). Finally, this new set of data, where we can measure/fit $\sigma(T)$ to a higher accuracy, lead to the prediction of the thermal runaway temperature that matches the measured onset flash temperature to a higher precision: specifically, the predicted $T_F$ for a thermal runaway is 598 °C *vs.* the measured $T_F$ for the onset of flash is 599 °C (Fig. 3(b)).

The above agreements (Fig. 3(b)) suggest that the flash sintering in ZnO powder specimens also starts as a result of thermal runaway (similar to the flash of ZnO single crystals) and the model proposed in §3 is applicable. We further note that a similar thermal runaway model proposed by Todd *et al.* [38] independently can also quantitatively explain the occurrence of flash in 3YSZ, an ionic conductor and the most extensively-studied flash sintering system, implying the general applicability of the thermal runaway mechanism for the start of flash beyond ZnO (an electronic conductor or a semiconductor self-doped with Zn interstadials and/or oxygen vacancies).

The specimen temperatures at the activated state have also been estimated based on Eq. (6). The measured conductivities for the activated state are higher than those extrapolated from a simple Arrhenius relation using the low-temperature data measured before the sintering; however, a simple Arrhenius relation is not expected for this case because the relative densities of the specimens increased from ~63.5% to ~90% after the flash sintering, accompanying with grain growth, which should affect (increase) conductivity in a nonlinear fashion that is difficult to predict. Alternatively, we estimated the conductivities based on the estimated specimen temperatures and an empirical relation obtained for polycrystalline ZnO specimens in prior studies [40, 41]; these estimated conductivities are listed in Table I, along with the measured conductivities. Specifically, the estimated specimen temperature is 1002 °C for the specimen with $I_{max}$ = 1A, which is more than 400 °C higher than the furnace temperature (~565 °C); the corresponding estimated conductivity at 1002 °C based on Ref. [40] is $8.4 \times 10^{-2}$ S/cm, which is close to the actual measured conductivity of $6.9 \times 10^{-2}$ S/cm. For the specimen with $I_{max}$ = 4A, the estimated specimen temperature is 1407 °C, more than 850 °C higher than the furnace temperature (~553 °C); the corresponding estimated conductivity at 1407 °C is $3.9 \times 10^{-2}$ S/cm, which is again close to the measured conductivity of $3.1 \times 10^{-2}$ S/cm. In both cases, the differences are only ~20% (even if the prior study measured ZnO polycrystal specimens with different/unknown microstructures [40]), suggesting that the conductivities at the activated state during flash sintering are similar to those of normal ZnO polycrystal specimens; thus, the presence of a substantial amount of non-equilibrium defects are not needed to explain the observed conductivities, though some non-equilibrium defects may exist and accelerate sintering rates and affect microstructural development

When the maximum current was set to be 1 A, a relative density of 90.8 % was achieved after the flash sintering (30 seconds in the activated state, where $J_{max} \approx 3.9$ A/cm$^2$ and $E \approx 52$ V/cm at the final steady state; noting that $E$ was set to 300 V/cm before the power source switched to the current-limited mode for all cases). When the maximum current was set to be 4 A ($J_{max} \approx 15.4$ A/cm$^2$ and $E \approx 44$ V/cm at the final steady state), a slightly lower sintered density (87.3%) was achieved. The reduction of densification at the high current density is presumably caused by excess grain growth. The measured grain size is >10 times greater and highly non-uniform in the latter case, which will be discussed in details in the next section. The results of sintered densities and measured grain sizes are summarized in Table I.

**4.3. Asymmetrical Microstructural Development: Potential-Induced Abnormal**

**Grain Growth**

    A particular intriguing and interesting observation of this study is represented by the anode-side abnormal grain growth and/or coarsening during the flash sintering of pure ZnO powder specimens; the measured grain sizes are substantially greater in anode (+) side than those at the cathode (-) sides. When the maximum current was set to be 1 A ($J_{max} \approx 3.9$ A/cm$^2$), the average grain sizes of cathode (-) and anode (+) sides, respectively, were measured to be $0.4 \pm {<}0.1$ μm and $0.9 \pm 0.1$ μm, respectively; this grain size disparity is clearly evident in Figs. 4(a) *vs.* 4(b) and Fig. 5. Since the starting particle size was slightly smaller than 0.5 μm, essentially no grain growth occurred at the cathode side, while the grain size doubled at the anode size. When the maximum current was set to be 4 A ($J_{max} \approx 15.4$ A/cm$^2$), the disparity increased further. The average grain sizes of the cathode and anode sides, respectively, were measured to be $3.5 \pm 1.8$ μm and $32.3 \pm 5.6$ μm, respectively (Fig. 6). In this case, there was substantial grain growth at the cathode size (by approximately 8×) and excess grain growth at the anode side (by approximately 80×); in the final state, the grains are about 10× larger at the anode side. In the most recent study of flash sintering of ZnO, AC fields (up to 160 V/cm) were used so that the asymmetric microstructural development and abnormal grain growth were not observed [37].

    The observation of enhanced grain growth at the anode side is somewhat in contrast to a prior report, where the grain growth was found to be enhanced at the cathode side in 8YSZ [34]. In that study, the grain growth occurred at the cathode side, which should not occur at that temperature without an electric current normally; Chen and co-workers attributed enhanced grain boundary mobility to the interaction of supersaturated oxygen vacancies and grain boundaries and a possible grain boundary reduction reaction that lowers the cation migration barriers [34]. Adapting and extending Chen and coworkers' theory from 8YSZ to ZnO, we propose that in the current case of ZnO, electrons accumulate at the anode size (due to the positive electric potential) and interact with surfaces and/or grain boundaries to enhance the interfacial transport rates via an oxidation reaction that increases the local cation vacancy concentration, which subsequently accelerates the coarsening of particles and/or grain growth during the flash sintering. Consistently, Tuller suggested that cations diffuse along ZnO grain boundaries via cation vacancies formed via an oxidation reaction at grain boundaries [35]. Adapting the grain boundary oxidation reaction originally proposed by Tuller [35], we propose that the following

defect chemical reactions may occur at ZnO grain boundaries in the anode side, induced by the presence of excess electrons:

$$e' + \tfrac{1}{2}O_2(\text{gas}) \rightarrow O_O^x + V_{Zn}'$$
$$e' + V_{Zn}' \rightarrow V_{Zn}''$$
(8)

Tuller also pointed out that the formation of Zn vacancies at grain boundaries appears to be counterintuitive since ZnO is normally believed to be a metal excess (or oxygen deficient) material [35, 39]. However, this disparity between the bulk and grain boundary defect structures can be well rationalized since the thermodynamic states [36, 42, 43] and (therefore) the defect structures [44, 45] of grain boundaries can often differ markedly from those of the corresponding bulk materials. If the above hypothesis of forming cation vacancies at ZnO grain boundaries (and free surfaces) induced by the positive electric potential and the corresponding accumulation of electrons is true, the grain growth and/or coarsening can be enhanced at the anode side, corroborating with the experimental observations made in the current study (Figs. 4-6).

Moreover, Fig. 5 shows an abrupt (discontinuous) transition between the large and small grains and this transitional line is located at ~56 μm away from the anode edge of the pure ZnO sample that was flash-sintered with $I_{\max} = 1$ A (noting that that the abnormal grain growth only occurred within a short distance to the anode in this specimen). A similar discontinuous transition was observed in 8YSZ by Kim *et al.* [34]. A discontinuous increase in grain boundary mobility and the associated abnormal grain growth indicate the possible occurrence of a grain boundary structural (complexion) transition [36], which, in the current case, is presumably a subtle structural transition associated with the formation of excess cation vacancies at grain boundaries (Eq. (8)) (that is difficult to be verified directly via experiments). Similar abnormal grain growth behaviors have been previously attributed to the formation of intergranular films [46] and complexion transitions [47-49] in two separate studies.

This abrupt transition (Fig. 5) also suggests that the disparity in grain growth is unlikely a simple result of temperature non-uniformity (in addition, the geometrical and heat transfer conditions were kept symmetric with respect to the anode and the cathode in our experiments); otherwise, a gradual change in the grain sizes would generally be expected (unless there is a temperature-induced discontinuous interfacial structural transition coincidentally).

## 4.4. Asymmetrical Microstructural Development: Growth of Single-Crystalline Rods against the Direction of the Electric Field

At the high current density of $J_{max} \approx 15.4$ A/cm$^2$ ($I_{max} = 4$ A), the growth of aligned single-crystalline rods and fibers was observed in the pure ZnO powder specimen, which presumably occurred at the cracks. Fig. 7(a) shows the growth of an array of such rods at ~1.4 mm away from the anode, and Fig. 7(b) shows some fibers grown at another location in an expanded view. These fibers and rods ranged from ~5 to >30 µm in their lengths. Many rods and fibers are hexagonal, and Fig. 7(c) shows enlarged views of some hexagonal rods. This observation suggests a directional growth of ZnO rods and fibers along the $c$ axis of the ZnO crystal structure; this directional growth along the <0001> direction is likely related to the polarity of the ZnO crystal structure, in which positively-charged Zn-(0001) planes and negatively-charged O-(000$\bar{1}$) planes are packed alternatingly [50].

It should be pointed out that the growth of ZnO rods may be related to local melting at cracks (therefore liquid-phase growth locally). Nonetheless, it is interesting to note that the alignment of ZnO rods does show an electric field effect. We should note that this experiment was performed at a current density that is higher than those commonly used in normal flash sintering. Nonetheless, the growth of aligned ZnO rods represents yet another interesting observation.

It is also interesting to note that all ZnO fibers and rods grew towards to the anode direction (Fig. 7). In comparison, Chen and co-workers found that pores migrated against the direction of the electric field via an ionomigration mechanism, along with enhanced grain growth in the cathode side, in a series of systematic studies of 8YSZ [25-27, 34]. In the current case, the ZnO rods and fibers grew against the direction of the electric field (Fig. 7), while the coarsening and grain growth were enhanced at the anode side (Figs. 4-6).

## 4.5. The Effects of Bi$_2$O$_3$ Doping

We further conducted experiments to test the effects of Bi$_2$O$_3$ doping on the flash sintering and microstructural development, which had not been conducted before, to reveal several interesting phenomena. First, addition of minor (0.5 mol. %) amount of Bi$_2$O$_3$ dopants defers the onset of the flash sintering of ZnO powder specimens from ($T_F =$) 553-565 °C to ($T_F =$) 620-621 °C, despite that Bi$_2$O$_3$ is typically considered as a sintering aid that should lower the sintering temperature of ZnO [51]. Moreover, this observation appears to be in contrast to several prior studies, where doping enabled or promoted the occurrence of flash sintering in $Al_2O_3$ (with MgO doping) [13] and $SnO_2$ (with $MnO_2$ doping) [14]. However, the observation of deferring onset flash sintering of ZnO by $Bi_2O_3$ doping can be readily understood via the doping effects on forming potential barriers at grain boundaries. $Bi_2O_3$-doped ZnO is a prototype system for varistors, where double Schottky barriers (as a result of space charges) form at grain boundaries, leading to nonlinear I-V behaviors [52]. The formation of such potential barriers at grain boundaries would likely reduce the conductivity of the powder specimens and defer the onset of the flash sintering.

The ZnO-$Bi_2O_3$ binary system has a eutectic reaction at 740 °C, above which a $Bi_2O_3$-based liquid phase forms; moreover, a series of prior studies demonstrated that nanometer-thick, premelting-like (quasi-liquid) films (that are 2-D interfacial "phases" and also called "complexions" [36, 42, 43, 47, 53-58]) can develop at both grain boundaries [51, 59, 60] and free surfaces [61-64] below the bulk eutectic temperature of 740 °C. All these facts suggest a modified flash sintering mechanism in $Bi_2O_3$-doped ZnO, where $\sigma(T)$ can no longer be represented by a simple Arrhenius relation (as being assumed for pure ZnO single crystals and powder specimens). Experimentally, the onset flash sintering occurred at the furnace temperatures of $T_F$ = 620-621 °C in the current study; the corresponding specimen temperatures are estimated to be 667-734 °C (Table I), which are slightly below the bulk eutectic temperature of 740 °C. This suggests that the occurrence of flash sintering is resulted from the formation of either a small amount of $Bi_2O_3$-based bulk liquid phase or interfacial liquid-like films (complexions), which changes (increases) the conductivity of the specimens substantially. This hypothesis is consistent with and supported by the measured conductivities shown in Fig. 2.

At the activated (flash sintering) state, the estimated specimen temperatures are 877 °C (for $J_{max} \approx 3.9$ A/cm$^2$) and 1254 °C (for $J_{max} \approx 15.2$ A/cm$^2$), respectively (Table I), well exceeding the eutectic temperature of 740 °C; thus, this is a case of liquid-phase sintering, which may be modified (enhanced) by applied electric currents.

Finally, $Bi_2O_3$ doping also made the grain growth and microstructural development uniform at both the cathode and anode sides. Fig. 8 (a) and (b) show the microstructures of cathode and anode sides of the 0.5 mol. % $Bi_2O_3$-doped ZnO specimen that was flash-sintered with $J_{max} \approx 3.9$ $A/cm^2$ (under the conditions identical to those used for the pure ZnO powder specimens); the measured grain sizes of cathode (-) and anode (+) sides, respectively, are $2.0 \pm 0.2$ μm and $1.7 \pm 0.2$ μm, respectively, which are essentially identical within the experimental errors. Furthermore, Fig. 8 shows a typical liquid-phase sintering microstructure with pores of ~0.5 μm in diameters being trapped at grain boundaries. A sintered density of 91.5 % has been achieved at this sintering condition ($J_{max} \approx 3.9$ $A/cm^2$), which is slightly higher than that of pure ZnO powder specimen flash-sintered at the identical conditions (Table I). At a high current density of $J_{max} \approx 15.2$ $A/cm^2$, the sintered density reduced to 88.2% of the theoretically density, which could again be attributed to excess particle coarsening and grain growth during the initial stage of flash sintering; the measured final grain size are ~13 μm, being identical at the cathode and anode sides within the experimental errors (Table I). The homogenization of microstructures and inhibition of the anode-side abnormal grain growth (that was observed for pure ZnO) can be attributed to a liquid-phase sintering effect.

## 5. Conclusions

DC flash sintering experiments were conducted for both pure and 0.5 mol. % $Bi_2O_3$-doped ZnO, along with benchmark flash experiments of ZnO single crystals, at a relatively high fixed initial applied electric field of 300 V/cm. The main findings and conclusions are summarized as follows:

- A key contribution of this study is the development of a model to predict the thermal runaway conditions that are coincidental with the observed onset flash temperatures in all ZnO based specimens.
- The excellent agreements between the predicted thermal runaway temperatures and observed onset flash temperatures authenticate the key underlying hypothesis, *i.e.*, the flash starts as a thermal runaway for at least ZnO single crystals and powder specimens (as well as 3YSZ as demonstrated independently by Todd *et al.* [38]) without the need of introducing an avalanche of non-equilibrium defects, although some non-equilibrium defects may form after the onset of the flash to accelerate the sintering (to account for the fast densification rates [6, 13, 18, 23]) and affect microstructural development; the $Bi_2O_3$

- doping leads to the formation of a bulk eutectic liquid or liquid-like interfacial complexion that causes a discontinuous increase in conductivity to initiate the flash sintering in a similar mechanism.

- Compared with single crystals, the flash of ZnO power specimens occurs at a substantially lower temperature, indicating the important roles of surfaces and grain boundaries; in particular, the reduction of onset flash temperature is explained from the enhanced conduction along the surfaces of ZnO particles, and this mechanism is also supported by the conductivity measurements and the model-experimental comparison.

- One intriguing and interesting observation is represented by the enhanced coarsening and/or grain growth at the anode side during the flash sintering, in contrast to the enhanced grain growth at the cathode side that was previously reported for 8YSZ [34]; the observation of a discontinuous transition between small and large grains suggests the occurrence of abnormal grain growth/coarsening resulted from an interfacial (defect) structural transition [36]; this cathode-side abnormal grain growth can be explained from the electric-potential-induced accumulation of electrons and an associated oxidation reaction to form excess cation vacancies at ZnO grain boundaries that promote interfacial diffusion, following Tuller's theory of ZnO grain boundary defect chemistry [35], as well as extending and combining Chen and colleagues' original concept of potential-induced abnormal grain growth [34] and the idea of grain boundary complexion transitions [36].

- Aligned growth of single-crystalline ZnO rods and fibers towards the anode direction was observed at a high current density for the pure ZnO powder specimen.

- $Bi_2O_3$ doping defers the onset of flash sintering, which is in contrast with prior studies where doping promoted flash sintering [13, 14]; however, this observation can be readily explained by the formation double Schottky barriers at grain boundaries of $Bi_2O_3$-doped ZnO, a prototype varistor material.

- $Bi_2O_3$ doping also homogenizes the microstructure (by eliminating the anode-side abnormal grain growth) via a liquid-phase sintering effect.

In summary, this study elucidates the flash mechanism, establishes a model to predict the flash conditions, and reports diversifying phenomena of sintering and grain growth under the influences of electric fields.

**Acknowledgement:**

This work is primarily supported by the Aerospace Materials for Extreme Environments program of the U.S. Air Force Office of Scientific Research (AFOSR) under the grants No. FA9550-10-1-0185 (05/2010 – 04/2013) and No. FA9550-14-1-0174 (09/2014 – 08/2019), and we thank our AFOSR program manager, Dr. Ali Sayir, for his support and guidance.  We also gratefully acknowledge partial supports from ONR (N00014-11-1-0678) and NSF (CMMI-1436305) during the AFOSR funding gap in 2013-2014.

**Table I:** Summary of key experimental results. The maximum current densities ($J_{max}$) were calculated based on the final specimen dimensions and measured electric currents. The initial electric field ($E$) was always set to be 300 V/cm; the power source switched to the current-limited mode for most specimens and the final electric fields ($E$) were calculated based the measured voltages and the final specimen thickness. The estimated conductivities for ZnO powder specimens were calculated based on the estimated specimen temperatures and an empirical relation of $\sigma(T) = 54e^{-0.71eV/kT}$ S/cm, reported in a prior study [40] for ZnO polycrystals.

| Sample | $I_{max}$ (A) | Sintered Relative Density | Measured Mean Grain Size ± One Standard Deviation (μm) | | Thermal Runaway vs. Onset Flash T | | After the Flash (in the Activated State) | | | | |
|---|---|---|---|---|---|---|---|---|---|---|---|
| | | | Cathode (-) Side | Anode (+) Side | $T_F$ (°C) | Estimated $T_S$ (°C) | Estimated $T_S$ (°C) | $E$ (V/cm) | $J_{max}$ (A/cm²) | Conductivity (S/cm) | |
| | | | | | | | | | | Measured | Estimated |
| ZnO Single Crystals | 1 | - | - | - | 870 | 962 | 1270 | 300 | 4.0 | $1.4 \times 10^{-2}$ | - |
| | 1.45 | - | - | - | 877 | 967 | 1417 | 274 | 5.8 | $1.8 \times 10^{-2}$ | - |
| Pure ZnO (Powder) | 1 | 90.8% | 0.4 ± <0.1 | 0.9 ± 0.1 | 565 | 675 | 1002 | 52 | 3.9 | $6.9 \times 10^{-2}$ | $8.4 \times 10^{-2}$ |
| | 4 | 87.3% | 3.5 ± 1.8 | 32.3 ± 5.6 | 553 | 662 | 1407 | 44 | 15.4 | $3.3 \times 10^{-1}$ | $3.9 \times 10^{-1}$ |
| ZnO + 0.5 mol. % $Bi_2O_3$ (Powder) | 1 | 91.5% | 2.0 ± 0.2 | 1.7 ± 0.2 | 621 | 734 | 877 | 31 | 3.9 | $1.2 \times 10^{-2}$ | - |
| | 4 | 88.2% | 13.1 ± 1.6 | 12.6 ± 1.1 | 620 | 667 | 1254 | 28 | 15.2 | $5.1 \times 10^{-2}$ | - |

# Figure Captions:

**Figure 1.** Measured electric power dissipation *vs.* furnace temperature curves for the flash sintering of pure and 0.5 mol. % Bi$_2$O$_3$-doped ZnO powder specimens and ZnO single crystals. For each of the three cases, experiments were conducted for two specimens, where we set $I_{max} \leq$ 1A and $I_{max} \leq$ 4A, respectively (noting that the current did not reach 4 A for the single crystal specimen). The open red squares represent a new experiment where we used a high-purity ZnO powder and a high-resolution ammeter so that we can measure more data points before the occurrence of the flash event (to allow a more critical test of the proposed thermal runaway model). The higher onset flash sintering temperature was likely due to the presence of less impurities (dopants to increase conductivity of ZnO) in the new ZnO powder specimen.

**Figure 2. (a)** Measured conductivity *vs.* the reciprocal of the estimated specimen temperature curves. **(b)** Measured conductivity *vs.* the reciprocal temperature for the new ZnO powder specimen, indicating Eq. (7) is applicable for the conductivity of the ZnO specimen before the flash.

**Figure 3.** Computed differential heat generation and dissipation rates *vs.* specimen temperature curves for the pure ZnO **(a)** single crystals and **(b)** powder specimens, respectively. Each panel includes data from two or three specimens. The flash condition (Eq. (2)) can be determined graphically by finding the intersection of the heat generation and dissipation rates curves, above which more heat is generated than that can be dissipated, leading to thermal runaway.

**Figure 4.** SEM micrographs of **(a)** the cathode (-) side and **(b)** the anode (+) side of a fractured surface of a pure ZnO specimen flash-sintered with a low current density.

**Figure 5.** SEM micrograph of a fractured surface of a pure ZnO specimen flash-sintered with a low current density, showing an abrupt transition from small to large grains. This transition occurred at a region that was located at about 56 μm away from the anode edge.

**Figure 6.** SEM micrographs of **(a)** the cathode (-) side and **(b)** the anode (+) side of a fractured surface of a pure ZnO specimen flash-sintered with a high current density.

**Figure 7. (a)** SEM micrographs of crystal rods grown at about 1.4 mm away from the anode edge of a pure ZnO specimen flash-sintered with a high current density. **(b)** An enlarged image of ZnO rods grown at a different region. **(c)** Enlarged views of some hexagonal rods. The growth of ZnO rods may be related to local melting at cracks (at a current density that is higher than those commonly used for normal flash sintering).

**Figure 8.** SEM micrographs of **(a)** the cathode (-) side and **(b)** the anode (+) side of a fractured surfaces of a 0.5 mol. % $Bi_2O_3$-doped ZnO specimen flash-sintered with a low current density.

## References:

[1] Z.A. Munir, U. Anselmi-Tamburini, M. Ohyanagi, The effect of electric field and pressure on the synthesis and consolidation of materials: A review of the spark plasma sintering method, J. Mater. Sci. 41 (2006) 763-777.

[2] Z.A. Munir, D.V. Quach, M. Ohyanagi, Electric current activation of sintering: A review of the pulsed electric current sintering process, J. Am. Ceram. Soc. 94 (2011) 1-19.

[3] M. Omori, Sintering, consolidation, reaction and crystal growth by the spark plasma system (sps), Materials Science and Engineering a-Structural Materials Properties Microstructure and Processing 287 (2000) 183-188.

[4] J.E. Garay, Current-activated, pressure-assisted densification of materials. In: Clarke DR, Ruhle M, Zok F, editors. Annual review of materials research, vol 40, vol. 40. 2010. p.445-468.

[5] R. Orru, R. Licheri, A.M. Locci, A. Cincotti, G.C. Cao, Consolidation/synthesis of materials by electric current activated/assisted sintering, Materials Science & Engineering R-Reports 63 (2009) 127-287.

[6] R. Raj, M. Cologna, J.S.C. Francis, Influence of externally imposed and internally generated electrical fields on grain growth, diffusional creep, sintering and related phenomena in ceramics, J. Am. Ceram. Soc. 94 (2011) 1941-1965.

[7] M. Cologna, B. Rashkova, R. Raj, Flash sintering of nanograin zirconia in < 5 s at 850 degrees c, J. Am. Ceram. Soc. 93 (2010) 3556-3559.

[8] M. Cologna, A.L.G. Prette, R. Raj, Flash-sintering of cubic yttria-stabilized zirconia at 750 degrees c for possible use in sofc manufacturing, J. Am. Ceram. Soc. 94 (2011) 316-319.

[9] A.L.G. Prette, M. Cologna, V. Sglavo, R. Raj, Flash-sintering of co2mno4 spinet for solid oxide fuel cell applications, J. Power Sources 196 (2011) 2061-2065.

[10] A. Gaur, V.M. Sglavo, Flash-sintering of mnco2o4 and its relation to phase stability, J. Eur. Ceram. Soc. 34 (2014) 2391-2400.

[11] A. Gaur, V.M. Sglavo, Densification of la0.6sr0.4co0.2fe0.8o3 ceramic by flash sintering at temperature less than 100 a degrees c, J. Mater. Sci. 49 (2014) 6321-6332.

[12] A. Karakuscu, M. Cologna, D. Yarotski, J. Won, J.S.C. Francis, R. Raj, B.P. Uberuaga, Defect structure of flash-sintered strontium titanate, J. Am. Ceram. Soc. 95 (2012) 2531-2536.

[13] M. Cologna, J.S.C. Francis, R. Raj, Field assisted and flash sintering of alumina and its relationship to conductivity and mgo-doping, J. Eur. Ceram. Soc. 31 (2011) 2827-2837.

[14] R. Muccillo, E.N.S. Muccillo, Electric field-assisted flash sintering of tin dioxide, J. Eur. Ceram. Soc. 34 (2014) 915-923.

[15] E. Zapata-Solvas, S. Bonilla, P.R. Wilshaw, R.I. Todd, Preliminary investigation of flash sintering of sic, J. Eur. Ceram. Soc. 33 (2013) 2811-2816.

[16] S.K. Jha, R. Raj, The effect of electric field on sintering and electrical conductivity of titania, J. Am. Ceram. Soc. 97 (2014) 527-534.

[17] X. Hao, Y. Liu, Z. Wang, J. Qiao, K. Sun, A novel sintering method to obtain fully dense gadolinia doped ceria by applying a direct current, J. Power Sources 210 (2012) 86-91.

[18] H. Yoshida, Y. Sakka, T. Yamamoto, J.-M. Lebrun, R. Raj, Densification behaviour and microstructural development in undoped yttria prepared by flash-sintering, J. Eur. Ceram. Soc. 34 (2014) 991-1000.

[19] J.S.C. Francis, R. Raj, Flash-sinterforging of nanograin zirconia: Field assisted sintering and superplasticity, J. Am. Ceram. Soc. 95 (2012) 138-146.

[20] R. Muccillo, M. Kleitz, E.N.S. Muccillo, Flash grain welding in yttria stabilized zirconia, J. Eur. Ceram. Soc. 31 (2011) 1517-1521.

[21] R. Muccillo, E.N.S. Muccillo, M. Kleitz, Densification and enhancement of the grain boundary conductivity of gadolinium-doped barium cerate by ultra fast flash grain welding, J. Eur. Ceram. Soc. 32 (2012) 2311-2316.

[22] R. Baraki, S. Schwarz, O. Guillon, Effect of electrical field/current on sintering of fully stabilized zirconia, J. Am. Ceram. Soc. 95 (2012) 75-78.

[23] R. Raj, Joule heating during flash-sintering, J. Eur. Ceram. Soc. 32 (2012) 2293-2301.

[24] J. Narayan, A new mechanism for field-assisted processing and flash sintering of materials, Scr. Mater. 69 (2013) 107-111.

[25] S.-W. Kim, S.-J.L. Kang, I.W. Chen, Ionomigration of pores and gas bubbles in yttria-stabilized cubic zirconia, J. Am. Ceram. Soc. 96 (2013) 1090-1098.

[26] S.-W. Kim, S.-J.L. Kang, I.W. Chen, Electro-sintering of yttria-stabilized cubic zirconia, J. Am. Ceram. Soc. 96 (2013) 1398-1406.

[27] I.W. Chen, S.-W. Kim, J. Li, S.-J.L. Kang, F. Huang, Ionomigration of neutral phases in ionic conductors, Advanced Energy Materials 2 (2012) 1383-1389.

[28] H. Conrad, D. Yang, Dependence of the sintering rate and related grain size of yttria-stabilized polycrystalline zirconia (3y-tzp) on the strength of an applied dc electric field, Materials Science and Engineering a-Structural Materials Properties Microstructure and Processing 528 (2011) 8523-8529.

[29] D. Yang, H. Conrad, Enhanced sintering rate and finer grain size in yttria-stablized zirconia (3y-tzp) with combined dc electric field and increased heating rate, Materials Science and Engineering a-Structural Materials Properties Microstructure and Processing 528 (2011) 1221-1225.

[30] J. Obare, W.D. Griffin, H. Conrad, Effects of heating rate and dc electric field during sintering on the grain size distribution in fully sintered tetragonal zirconia polycrystals stabilized with 3% molar yttria (3y-tzp), J. Mater. Sci. 47 (2012) 5141-5147.

[31] H. Conrad, Space charge and grain boundary energy in zirconia (3y-tzp), J. Am. Ceram. Soc. 94 (2011) 3641-3642.

[32] D. Yang, H. Conrad, Enhanced sintering rate of zirconia (3y-tzp) by application of a small ac electric field, Scr. Mater. 63 (2010) 328-331.

[33] S. Ghosh, A.H. Chokshi, P. Lee, R. Raj, A huge effect of weak dc electrical fields on grain growth in zirconia, J. Am. Ceram. Soc. 92 (2009) 1856-1859.

[34] S.W. Kim, S.G. Kim, J.I. Jung, S.J.L. Kang, I.W. Chen, Enhanced grain boundary mobility in yttria-stabilized cubic zirconia under an electric current, J. Am. Ceram. Soc. 94 (2011) 4231-4238.

[35] H.L. Tuller, Zno grain boundaries: Electrical activity and diffusion, Journal of Electroceramics 4 (1999) 33-40.

[36] P.R. Cantwell, M. Tang, S.J. Dillon, J. Luo, G.S. Rohrer, M.P. Harmer, Overview no. 152: Grain boundary complexions, Acta Mater. 62 (2014) 1-48.

[37] C. Schmerbauch, J. Gonzalez-Julian, R. Roeder, C. Ronning, O. Guillon, Flash sintering of nanocrystalline zinc oxide and its influence on microstructure and defect formation, J. Am. Ceram. Soc. 97 (2014) 1728-1735.

[38] R.I. Todd, E. Zapata-Solvas, R.S. Bonilla, T. Sneddon, P.R. Wilshaw, Electrical characteristics of flash sintering: Thermal runaway of joule heating, J. Eur. Ceram. Soc. 35 (2015) 1865-1877.

[39] M.D. McCluskey, S.J. Jokela, Defects in zno, J. Appl. Phys. 106 (2009) 071101.

[40] P.H. Miller, The electrical conductivity of zinc oxide, Physical Review 60 (1941) 890-895.

[41] J.P. Han, P.Q. Mantas, A.M.R. Senos, Defect chemistry and electrical characteristics of undoped and mn-doped zno, J. Eur. Ceram. Soc. 22 (2002) 49-59.

[42] J. Luo, Developing interfacial phase diagrams for applications in activated sintering and beyond: Current status and future directions, J. Am. Ceram. Soc. 95 (2012) 2358-2371.

[43] W.D. Kaplan, D. Chatain, P. Wynblatt, W.C. Carter, A review of wetting versus adsorption, complexions, and related phenomena: The rosetta stone of wetting J. Mater. Sci. 48 (2013) 5681-5717.

[44] Y.M. Chiang, E.B. Lavik, D.A. Blom, Defect thermodynamics and electrical properties of nanocrystalline oxides: Pure and doped ceo2, Nanostructured Materials 9 (1997) 633-642.

[45] Y.M. Chiang, E.B. Lavik, I. Kosacki, H.L. Tuller, J.Y. Ying, Defect and transport properties of nanocrystalline ceo2-x, Appl. Phys. Lett. 69 (1996) 185-187.

[46] I. MacLaren, R.M. Cannon, M.A. Gülgün, R. Voytovych, N.P. Pogrion, C. Scheu, U. Täffner, M. Rühle, Abnormal grain growth in alumina: Synergistic effects of yttria and silica, J. Am. Ceram. Soc. 86 (2003) 650.

[47] S.J. Dillon, M. Tang, W.C. Carter, M.P. Harmer, Complexion: A new concept for kinetic engineering in materials science, Acta Mater. 55 (2007) 6208-6218

[48] M.P. Harmer, Interfacial kinetic engineering: How far have we come since kingery's inaugural sosman address?, J. Am. Ceram. Soc. 93 (2010) 301-317.

[49] S.J. Dillon, M.P. Harmer, Demystifying the role of sintering additives with "complexion", J. Eur. Ceram. Soc. 28 (2008) 1485-1493.

[50] N.S. Ramgir, D.J. Late, A.B. Bhise, M.A. More, I.S. Mulla, D.S. Joag, K. Vijayamohanan, Zno multipods, submicron wires, and spherical structures and their unique field emission behavior, J. Phys. Chem. B 110 (2006) 18236-18242.

[51] J. Luo, H. Wang, Y.-M. Chiang, Origin of solid state activated sintering in bi2o3-doped zno, J. Am. Ceram. Soc. 82 (1999) 916.

[52] D.R. Clarke, Varistor ceramics, J. Am. Ceram. Soc. 82 (1999) 485-502.

[53] J. Luo, H. Cheng, K.M. Asl, C.J. Kiely, M.P. Harmer, The role of a bilayer interfacial phase on liquid metal embrittlement, Science 333 (2011) 1730-1733.

[54] M.P. Harmer, The phase behavior of interfaces, Science 332 (2011) 182-183.

[55] M. Tang, W.C. Carter, R.M. Cannon, Grain boundary transitions in binary alloys, Phys. Rev. Lett. 97 (2006) 075502.

[56] M. Tang, W.C. Carter, R.M. Cannon, Diffuse interface model for structural transitions of grain boundaries, Physical Review B 73 (2006) 024102.

[57] X. Shi, J. Luo, Developing grain boundary diagrams as a materials science tool: A case study of nickel-doped molybdenum, Physical Review B 84 (2011) 014105.

[58] M. Baram, D. Chatain, W.D. Kaplan, Nanometer-thick equilibrium films: The interface between thermodynmaics and atomistics, Science 332 (2011) 206-209.

[59] H. Wang, Y.-M. Chiang, Thermodynamic stability of intergranular amorphous films in bismuth-doped zinc oxide, J. Am. Ceram. Soc. 81 (1998) 89-96.

[60] Y.-M. Chiang, H. Wang, J.-R. Lee, Hrem and stem of intergranular films at zinc oxide varistor grain boundaries, J. Microsc. 191 (1998) 275-285.

[61] J. Luo, Y.-M. Chiang, Wetting and prewetting on ceramic surfaces, Annu. Rev. Mater. Res. 38 (2008) 227-249.

[62] J. Luo, Y.-M. Chiang, R.M. Cannon, Nanometer-thick surficial films in oxides as a case of prewetting, Langmuir 21 (2005) 7358-7365.

[63] J. Luo, Y.-M. Chiang, Existence and stability of nanometer-thick disordered films on oxide surfaces, Acta Mater. 48 (2000) 4501-4515.

[64] J. Luo, Y.-M. Chiang, Equilibrium-thickness amorphous films on {11-20} surfaces of $bi_2o_3$-doped zno, J. Eur. Ceram. Soc. 19 (1999) 697-701.

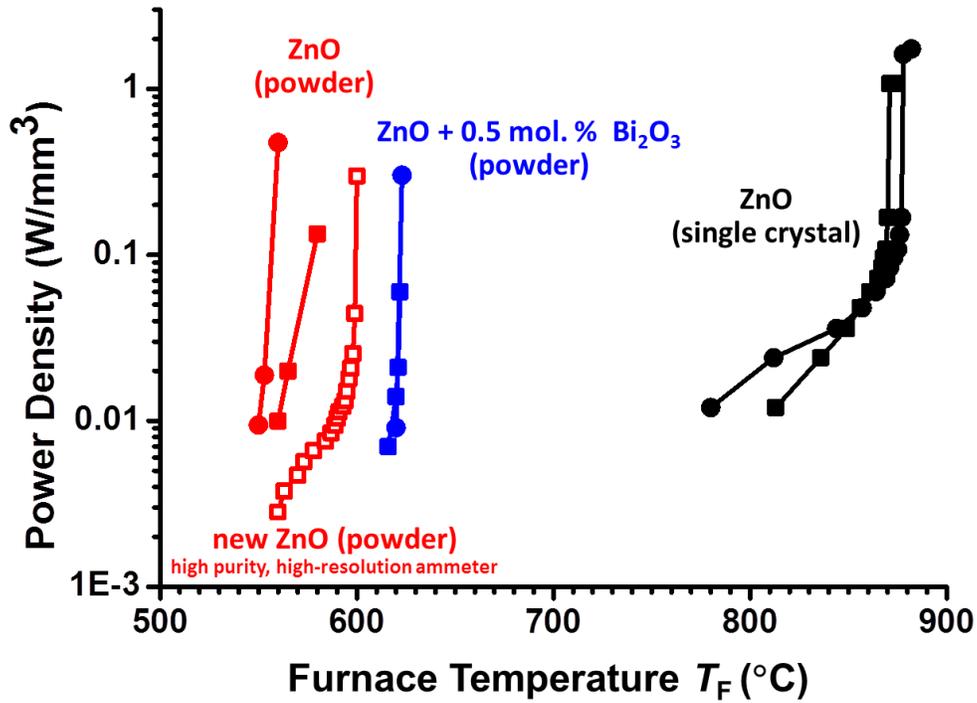

**Figure 1.** Measured electric power dissipation *vs.* furnace temperature curves for the flash sintering of pure and 0.5 mol. % $Bi_2O_3$-doped ZnO powder specimens and ZnO single crystals. For each of the three cases, experiments were conducted for two specimens, where we set $I_{max} \leq$ 1A and $I_{max} \leq$ 4A, respectively (noting that the current did not reach 4A for the single crystal specimen). The open red squares represent a new experiment where we used a high-purity ZnO powder and a high-resolution ammeter so that we can measure more data points before the occurrence of the flash event (to allow a more critical test of the proposed thermal runaway model). The higher onset flash sintering temperature was likely due to the presence of less impurities (dopants to increase conductivity of ZnO) in the new ZnO powder specimen.

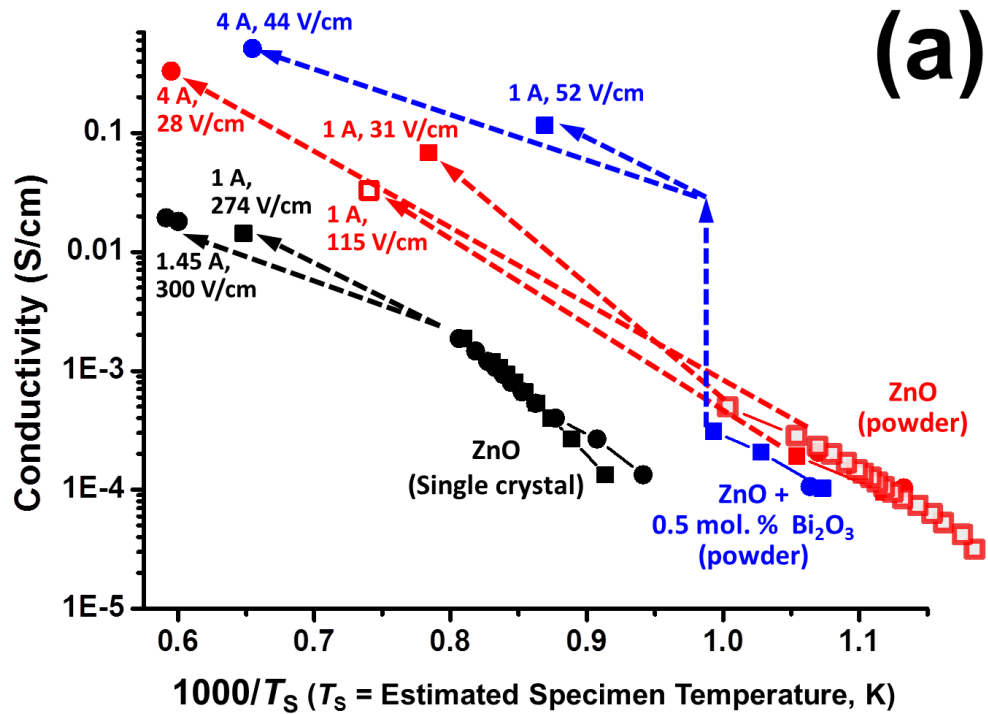

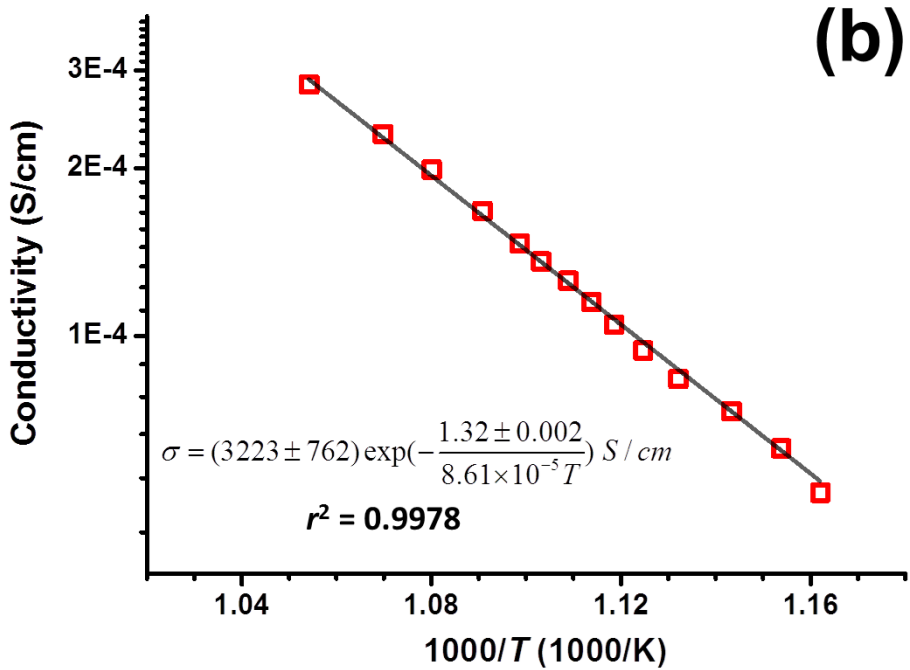

**Figure 2. (a)** Measured conductivity *vs.* the reciprocal of the estimated specimen temperature curves. **(b)** Measured conductivity *vs.* the reciprocal temperature for the new ZnO powder specimen, indicating Eq. (7) is applicable for the conductivity of the ZnO specimen before the flash.

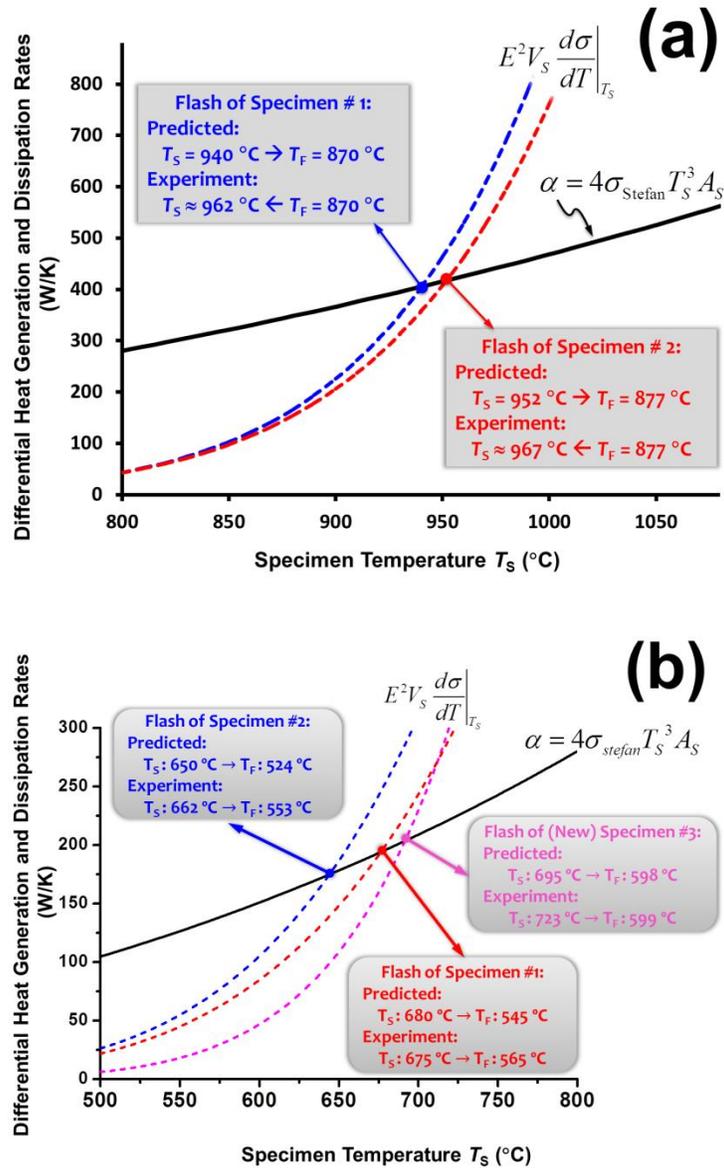

**Figure 3.** Computed differential heat generation and dissipation rates *vs.* specimen temperature curves for the pure ZnO **(a)** single crystals and **(b)** powder specimens, respectively. Each panel includes data from two or three specimens. The flash condition (Eq. (2)) can be determined graphically by finding the intersection of the heat generation and dissipation rates curves, above which more heat is generated than that can be dissipated, leading to thermal runaway.

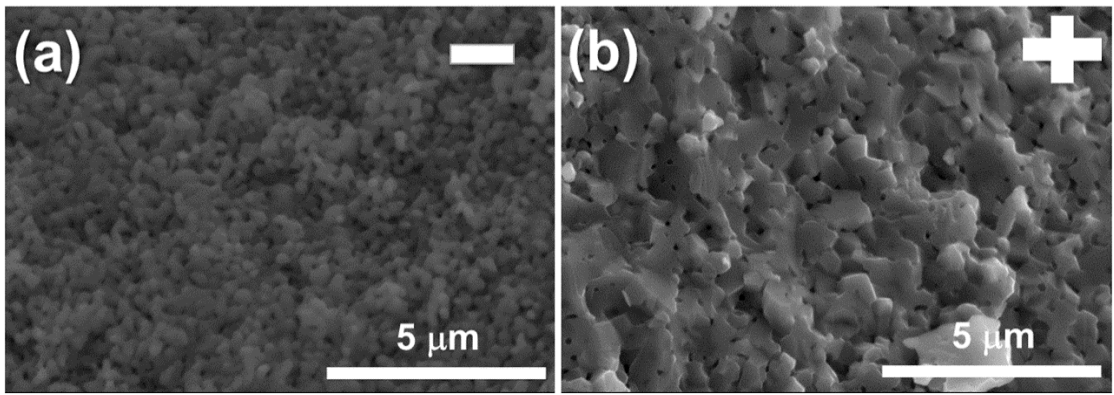

**Figure 4.** SEM micrographs of **(a)** the cathode (-) side and **(b)** the anode (+) side of a fractured surface of a pure ZnO specimen flash-sintered with a low current density.

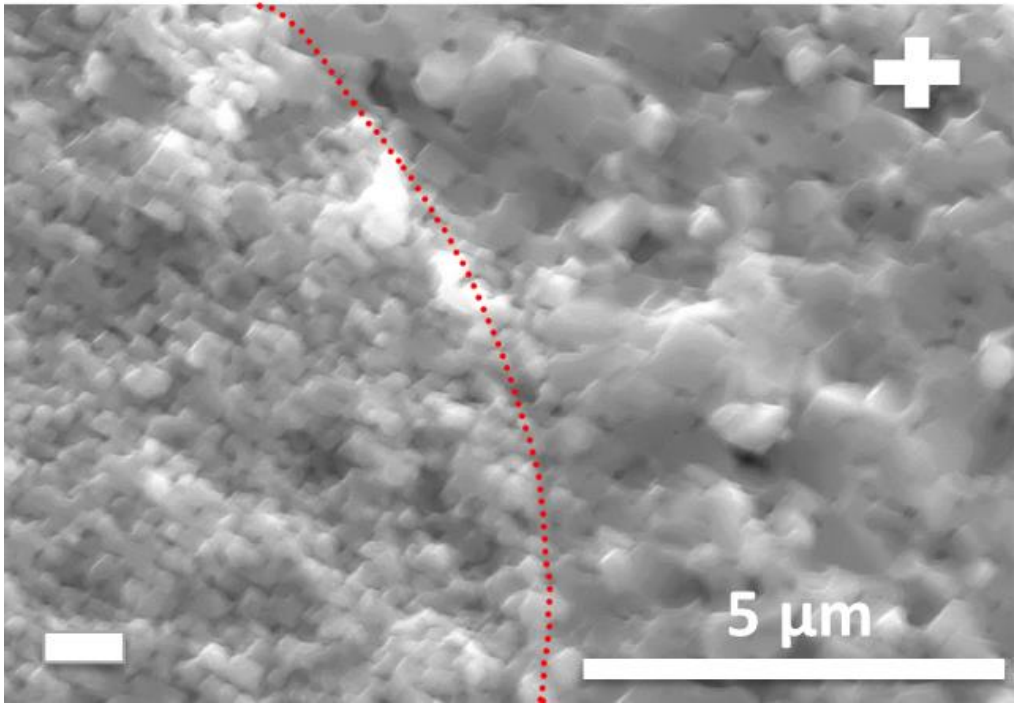

**Figure 5.** SEM micrograph of a fractured surface of a pure ZnO specimen flash-sintered with a low current density, showing an abrupt transition from small to large grains. This transition occurred at a region that was located at about 56 μm away from the anode edge.

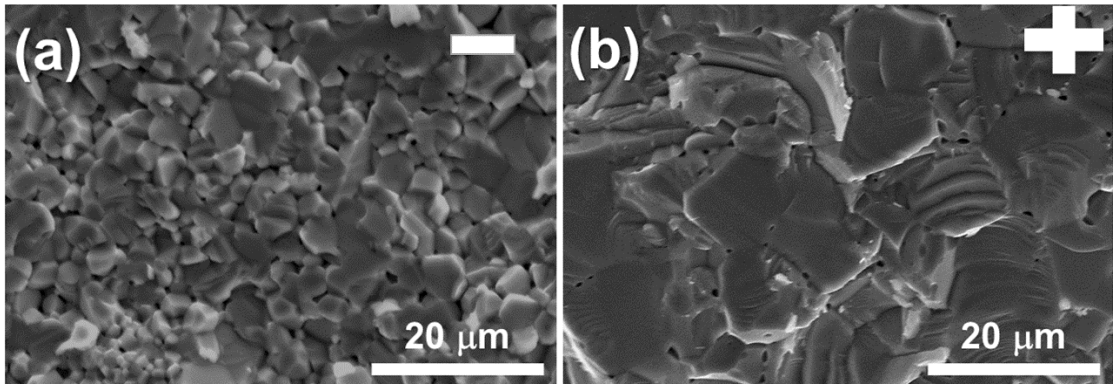

**Figure 6.** SEM micrographs of **(a)** the cathode (-) side and **(b)** the anode (+) side of a fractured surface of a pure ZnO specimen flash-sintered with a high current density.

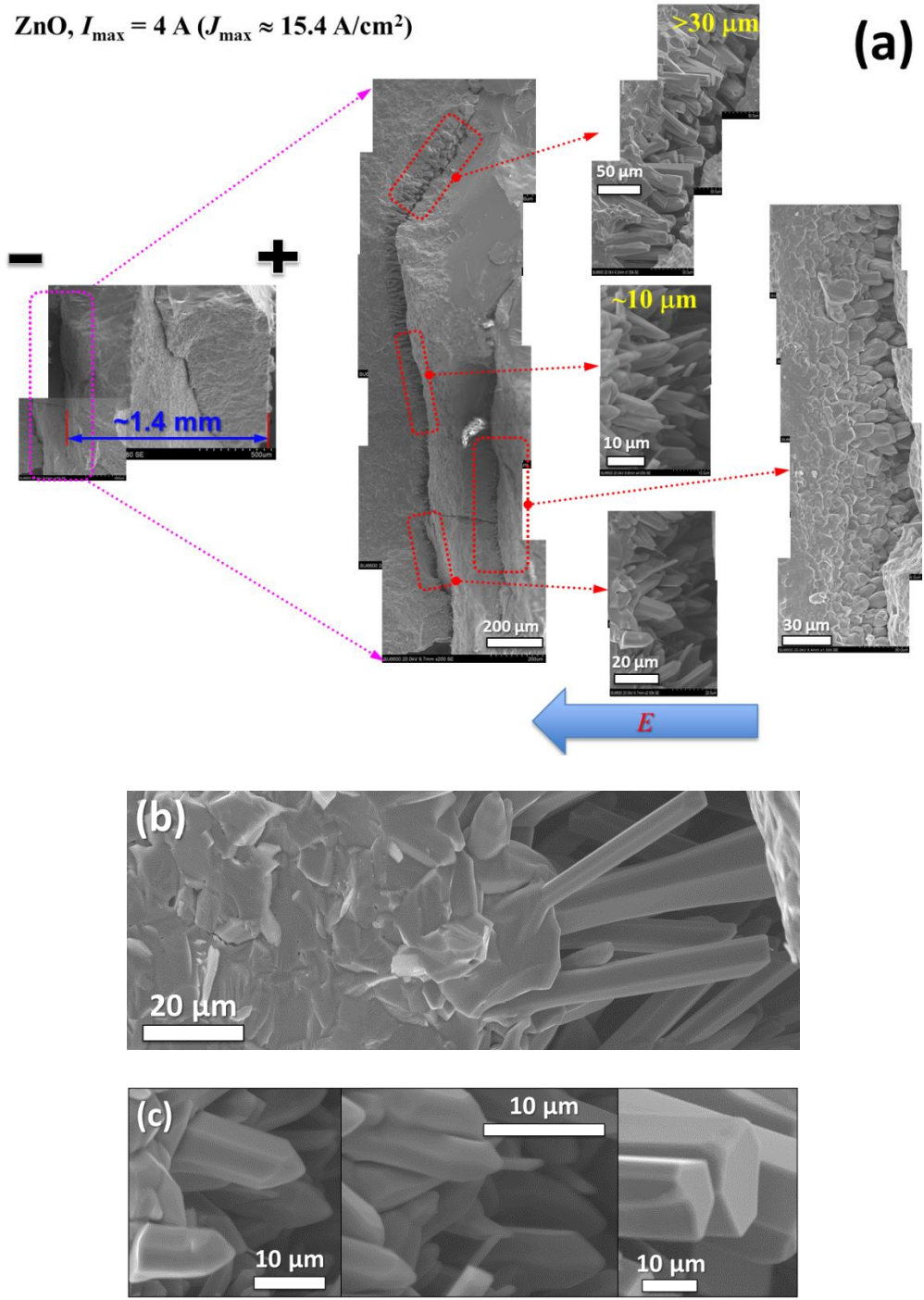

**Figure 7. (a)** SEM micrographs of crystal rods grown at about 1.4 mm away from the anode edge of a pure ZnO specimen flash-sintered with a high current density. **(b)** An enlarged image of the ZnO rods grown at a different region. **(c)** Enlarged views of some hexagonal rods. The growth of ZnO rods may be related to local melting at cracks (at a current density that is higher than those commonly used for normal flash sintering).

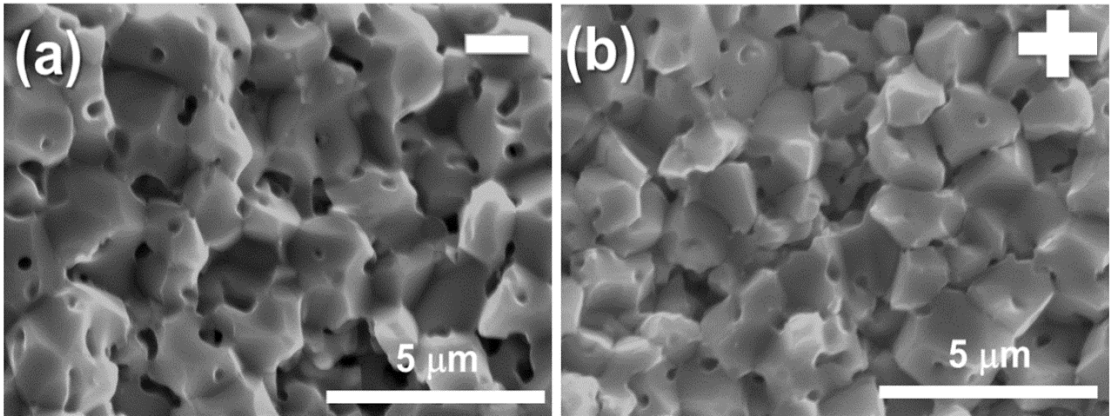

**Figure 8.** SEM micrographs of **(a)** the cathode (-) side and **(b)** the anode (+) side of a fractured surfaces of a 0.5 mol. % $Bi_2O_3$-doped ZnO specimen flash-sintered with a low current density.